\documentclass[%
 reprint,
 amsmath,amssymb,
 aps,
prb,
]{revtex4-2}

\usepackage{graphicx}
\usepackage{dcolumn}
\usepackage{bm}
\usepackage{xcolor}
\usepackage[version=4]{mhchem}

\usepackage{booktabs}

\begin{document}

\preprint{APS/123-QED}

\title{An Embedding Cluster Approach for Accurate Electronic Structure Calculations of (229)Th:\ce{CaF2}.}

\author{Kamil Nalikowski$^{1}$}
\author{Valera Veryazov$^{2}$}
\author{Kjeld Beeks$^{3,4}$}
\author{Thorsten Schumm$^{3}$}
\email{email: thorsten.schumm@tuwien.ac.at}
\author{Marek Kro\'{s}nicki$^{1}$}
\email{email: marek.krosnicki@ug.edu.pl}

\affiliation{$^1$Institute of Theoretical Physics and Astrophysics, Faculty of Mathematics Physics and Informatics, University of Gda\'{n}sk, ul. Wita Stwosza 57, Gda\'{n}sk, 80-308, Poland}
\affiliation{$^2$Division of Computational Chemistry, Chemical Center, Lund University, P.O. Box 124, SE-221 00 Lund, Sweden}
\affiliation{$^3$Vienna Center for Quantum Science and Technology, Atominstitut, TU Wien, 1020 Vienna, Austria}
\affiliation{$^4$Laboratory for Ultrafast Microscopy and Electron Scattering (LUMES), Institute of Physics, École Polytechnique Fédérale de Lausanne (EPFL), Lausanne CH-1015, Switzerland}

\date{\today}

\begin{abstract}
Building on recent advances of the embedded cluster approach combined with multiconfigurational theory, this work investigates the electronic states in thorium-doped \ce{CaF2} crystals. Th:\ce{CaF2} is currently establishing as a promising material for solid-state nuclear clocks, which utilize  the laser-accessible isomeric state in thorium-229.
By comparing simulated absorption spectra of a library of defect configurations with experimental data, we demonstrate
the impact of fluorine vacancies and calcium vacancies on the Th:\ce{CaF2} electronic structure. Our
results indicate that fluorine-deficient sites can introduce local electronic states within the band gap,
resonant with the isomer energy, potentially contributing to non-radiative decay or quenching of the
Th-229 isomer. We also explore the potential of electron-nuclear bridge mechanisms to enhance nuclear
excitation or de-excitation, offering a pathway for more efficient control over the nuclear clock. This
study provides key insights for optimizing the crystal environment for nuclear metrology applications
and opens new avenues for further experimental and theoretical exploration of thorium-doped
ionic crystals.

\end{abstract}

\maketitle


\textit{Introduction.}-- 
Thorium-229 has an unusually low-energy first nuclear excited state at 8.4\,eV, which has a long lifetime, therefore called an isomer~\cite{beeks2021thorium}. As 8.4\,eV is within reach of contemporary laser technology, nuclear and solid-state metrology applications have been proposed, where the largest emphasis is on developing a nuclear clock~\cite{tkalya1996processes,peik2003nuclear,tkalya2011proposal,yamaguchi2024laser,tiedau2024laser,elwell2024laser,zhang2024frequency}. The nuclear isomer energy exceeds the first ionization energy of neutral thorium atoms and thus the nucleus can decay either radiatively, under emission of a photon, or via internal conversion (IC), where energy is transferred from the nucleus to an atomic shell electron which is then ejected. The IC process has a dramatically reduced lifetime of the order of $\mu$s~\cite{seiferle2019energy} compared to the $\sim$2500\,s lifetime of the radiative decay in vacuum~\cite{tiedau2024laser,yamaguchi2024laser,pineda2024radiative}. For any metrology application, a long lifetime with a corresponding narrow laser resonance linewidth, is imperative.

To inhibit the IC process, thorium is either ionized and trapped electromagnetically or doped into an ionic crystal~\cite{peik2003nuclear}. The band gap of the host crystal should be large enough to ensure crystal transparency for the vacuum ultraviolet (VUV) light used to excite the isomer. Additionally, the local electronic states generated by the thorium impurity should not facilitate IC. However, local electronic states providing electronic dipole transitions could potentially be used to enhance the pumping of the isomer's hyperfine structure, exploiting electron-nuclear bridge (EB) processes \cite{nickerson2020nuclear,nickerson2021driven}. 

The \ce{CaF_2} crystal is transparent to VUV light, has a direct band gap of 11.8\,eV and was therefore used as a host material for Th~\cite{beeks2022nuclear,beeks2023growth}. 
The local environment of the Th impurity was studied using density functional theory (DFT) methods under periodic boundary conditions~\cite{dessovic2014229thorium,pimon2022ab}. The authors assumed a local charge compensation scheme and concluded, that in a pure ionic bond model, thorium is in a 4+ charge state, with the thorium site being compensated by two additional \ce{F^-} interstitials. They stated that other charge compensation configurations, such as a Ca vacancy, are less energetically favourable. 

From an experimental perspective however, several effects can inhibit reaching complete charge compensation for all, or even the majority, of \textsuperscript{229}Th atoms in the crystal. For crystal doping, radiolysis occurring during growth of \textsuperscript{229}Th:CaF\textsubscript{2} causes strong losses of fluorine, resulting in a non-stoichiometric crystal that lacks F compared to perfect CaF\textsubscript{2}. VUV spectral absorption measurements performed for different fluorine content showed that at low F content, absorption peaks are present around 124\,nm, 130\,nm and 145\,nm. At stoichiometric F content, only an absorption peak at 124\,nm is present~\cite{beeks2024optical}. In fluorescence, a peak at 180\,nm was observed~\cite{beeks2022nuclear}. In samples where \textsuperscript{229}Th is mechanically implanted into stoichiometric \ce{CaF_2} after growth, a fluorescence peak at 180\,nm was again observed~\cite{kraemer2023observation}, together with a broad fluorescence at 141\,nm~\cite{pineda2024radiative}.

This indicates that non-compensated \textsuperscript{229}Th defects occur in \ce{CaF_2} as a result of radioactivity during growth and/or implantation, which are associated with the emergence of excited electronic states. A theoretical description of \ce{Th} substitution in \ce{CaF_2} is important not only for characterization of the optical spectra and assigning the observed peaks to the specific electron transitions, but also to clarify the changes in the chemical bonding caused by replacement of Ca(II) atom with Th(IV).

\textit{Theoretical model of \ce{Th}:\ce{CaF_2}}.-- 
\ce{CaF_2} is known for irregularities in the crystal structure. Fluorine atoms can move in the lattice, creating an effect of ionic conductivity below the melting point (superionic phase)~\cite{VORONIN20011349}. The presence of \ce{Th} in the structure adds additional disturbance. The increased valency of \ce{Th} in comparison to \ce{Ca} can be compensated by different mechanisms, ranging from adding extra fluorine atoms, to the re-arranging of chemical bonds. 

\begin{figure}
    \centering
    \includegraphics[width=\linewidth]{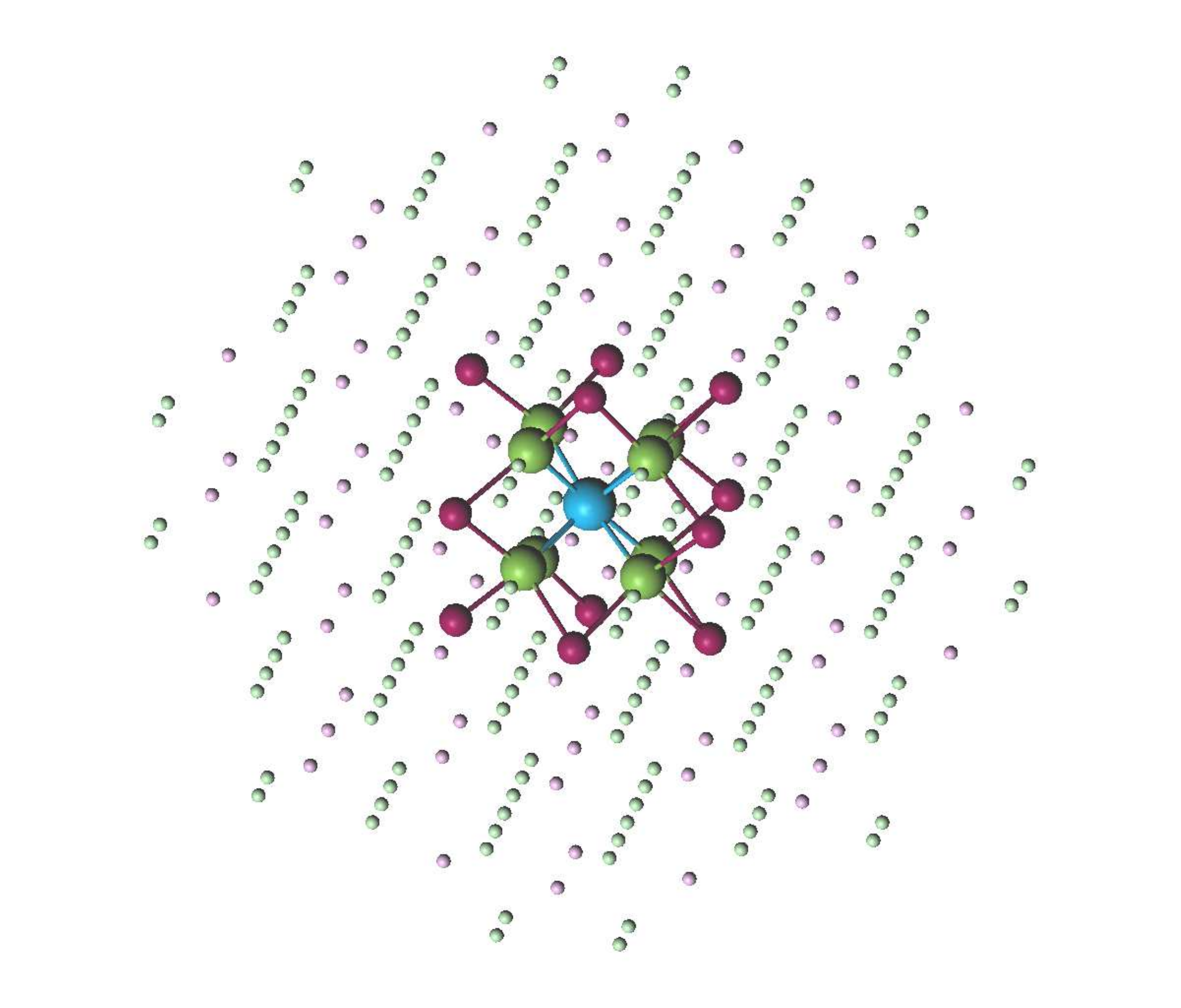}
    \caption{The Quantum Part of the embedded cluster  (\ce{Ca^{+2}} vacancy, \ce{[ThF_8Ca_11]^{+18}}), surrounded by model potentials. Point charges are not shown. Cyan - Thorium, Green - Fluorine, Red - Calcium 
    }
    \label{fig:f8ca11}
\end{figure}

A small part of a crystal, a cluster, can be considered as a "Quantum Part" of the system, where an approximate variant of the Schr\"odinger equation for the electrons can be solved. The selection of such a cluster around our point of interest, e.g. a point defect, results in a non-stoichiometric composition and thus a large formal charge. In order to compensate this charge, the cluster must be embedded into the electrostatic field of surrounding point charges. As a final step, a buffer between the Quantum Part and the surrounding charges should be introduced, for instance by implementing the model's atomic potentials (see Fig.~\ref{fig:f8ca11}). 

The cluster approach has two advantages in comparison to the periodic models: i) it is possible to model sparse point defects, including charged defects, ii) the description of the electronic structure is not limited by DFT, but may include more reliable methods, like post-Hartree-Fock methods, and in particular, multiconfigurational theory~\cite{Barandiaran2022}.

We selected several clusters (see Table~\ref{tab:properties}) to model different types of environment for the \ce{Th} dopant.
The clusters form the Quantum Part, which is treated by multi-configurational theory. All clusters were embedded into a shell of ab initio model potentials (AIMP)~\cite{Pascual2009, Krosnicki2014, DeGraaf1998, Larsson2022chemphys} and a set of point charges~\cite{Sushko2010}. Such embedding is necessary to provide the correct electrostatic potentials around the cluster and to compensate the formal charge of the cluster. For simplicity, we will refer to the Quantum Part (see Fig.~\ref{fig:f8ca11}) to denote the whole embedding in the following text.

The free neutral thorium atom has electron configuration [\ce{Rn}]$6d^27s^2$, and its usual oxidation state is 4+ when forming purely ionic compounds. However,  there are examples in the literature that thorium exhibits other oxidation states when forming fluoride compounds~\cite{doi:10.1021/acs.jpca.5b02544}.
To investigate the local properties of the electronic structure of thorium, we adapted the full valence definition~\cite{Valency} for the case of non-orthogonal atomic basis sets.

\textit{Computational details}.--
A theoretical description of the electronic structure of \ce{Th}:\ce{CaF_2} requires a computational technique which is capable of describing the ground and excited state with a high accuracy. An embedded cluster approach in combination with multiconfigurational theory (CASSCF/CASPT2 - Complete Active Space Self-Consistent Field with a second order perturbation)~\cite{Andersson1992,ThePinkBook} can be used for this purpose. Spin-orbit coupling was included by applying the CAS-SI (State Interaction) approach~\cite{Malmqvist1989}, which can mix several states (even with a different multiplicity), variationally optimize the resulting wavefunction and compute the transition moments between the states~\cite{Malmqvist2002}. 
Relativistic effects were treated by applying a DKH2 (Douglas-Kroll-Hess) Hamiltonian~\cite{dkh}, in connection to relativistic basis sets: ANO-RCC-VTZP for \ce{Th}~\cite{RCC} and ANO-XS basis set for other atoms~\cite{XS}.

All calculations of the electronic structure of embedded clusters were performed with the Molcas~8.6 code~\cite{Aquilante2020,openmolcas2023}. 

\begin{table*}[]
    \centering
    \begin{tabular}{|ll|cc|r|}
    \toprule \midrule
    Quantum Part&Site description & Th Charge &  Th Valence & Excitation  \\
       &&          & & Energy, eV \\ \midrule
         \ce{[ThF_6Ca_12]^{+22}}& double \ce{F^-} vacancy& 2.86 (2.14) & 4.18 (3.41)& 6.19 \\
         \ce{[ThF_7Ca_12]^{+21}}& single \ce{F^-} vacancy& 2.78 (2.05) & 4.20 (3.50)&  8.07\\
         \ce{[ThF_8Ca_11]^{+18}}& \ce{Ca^{+2}} vacancy& 2.69 (1.87) & 4.21 (3.36)&  10.02\\
         \ce{[ThF_8Ca_12]^{+20}}& 8 \ce{F-} coordinated Th(IV)& 2.73 (1.92) & 4.22 (3.44)& 10.39 \\
        \ce{[ThF_10Ca_12]^{+18}}& 10 \ce{F^-} coordinated Th(IV)& 2.71 (1.87) & 4.21 (3.39)& 11.35\\
         \midrule \bottomrule
    \end{tabular}

    \caption{
    Description of the clusters used in calculations, charge and valence of \ce{Th} atom and the CASPT2 energy of the first excited state in each Quantum Part. Values in parenthesis indicate parameters of the first excited state. }
    \label{tab:properties}
\end{table*}

\begin{figure}
    \centering
    \includegraphics[width=0.7\linewidth]{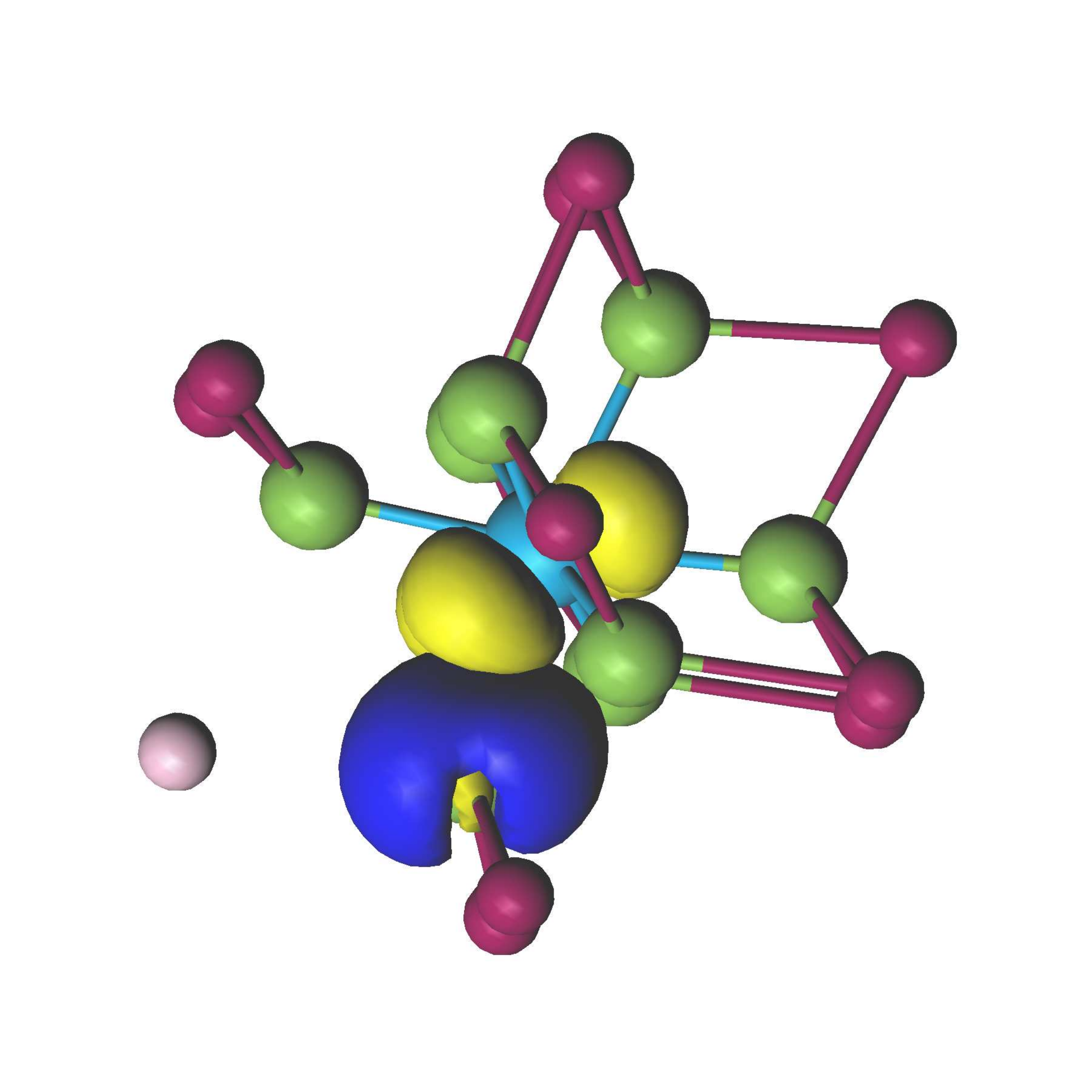}
    \includegraphics[width=0.7\linewidth]{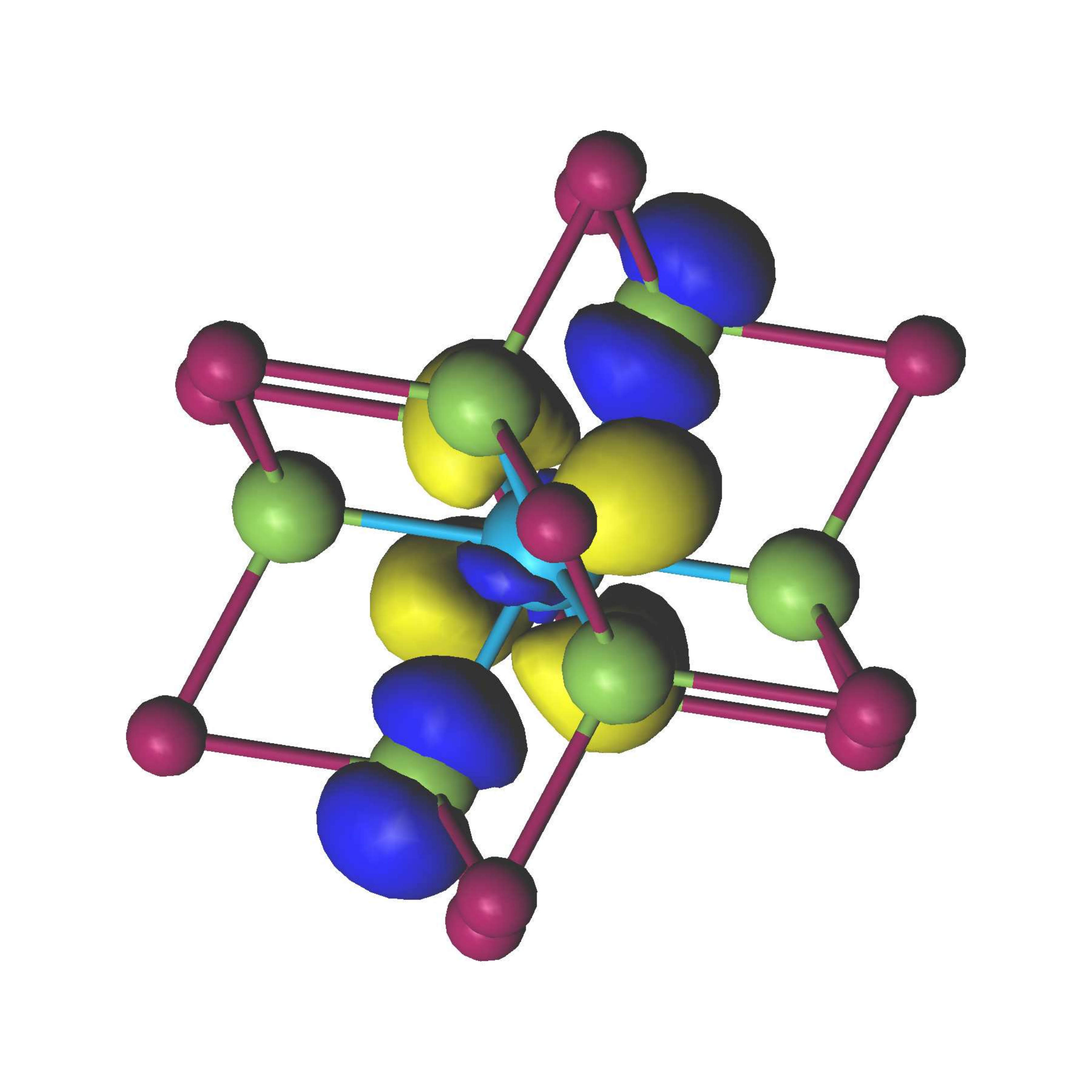}
    \caption{Electron density difference between the ground state and the first excited state for the Quantum Part region of \ce{[ThF_8Ca_11]^{+18}} (top) and \ce{[ThF_8Ca_12]^{+20}} (bottom) indicating charge transfer excitation from F to Th. A pale pink atom indicates the position of the Ca vacancy. The blue and yellow regions indicate areas where the electron density is lower and higher in the excited state compared to the ground state, respectively.}
    \label{fig:vCadiff}
\end{figure}

\textit{Results and discussion.}--
The results of the calculations of the electronic structure are summarized in Table~\ref{tab:properties}. 
Based on the analysis of the electronic density, we can conclude that for each of these clusters, thorium is four valent in its ground state, and the atomic bonds formed with fluorine have a mixed ionic-covalent character. This is an important result, as previously, a Th 4+ charge state was associated with the two "fully" charge compensated clusters \ce{[ThF_8Ca_11]^{+18}} and \ce{[ThF_10Ca_12]^{+18}} only. The lowest excited states arise from charge transfer from fluorine to thorium, resulting in thorium becoming three valent. This transition can also be observed by comparing the electron density, computed for the ground and the first excited state, as shown in Figure~\ref{fig:vCadiff}.

In contrast to DFT, the multiconfigurational approach can optimize both the ground and excited states, thus the excitation energies are computed as differences between total energies of different states. Table~\ref{tab:properties} presents a list of excitation energies, which are discussed below. 

On the other hand, the total electronic energy of a defect embedded in a crystal lattice cannot be easily decomposed, making it difficult to conclusively assess the enthalpy of a particular cluster chemical composition.
However, if one compares the calculated total energies of four clusters, which differ only by the amount of fluorine ion: \ce{[ThF_6Ca_{12}]^{+22}}, \ce{[ThF_7Ca_{12}]^{+21}}, \ce{[ThF_8Ca_{12}]^{+20}} and \ce{[ThF_{10}Ca_{12}]^{+18}}, the energy difference  per fluorine ion between subsequent clusters  (in a.u.) decreases as 100.5, 100.4, 99.9. while the energy of an \ce{F^-} ion with the same basis set is about 99.6 a.u. The energy trend per added fluorine ion suggests that the system \ce{[ThF_{10}Ca_{12}]^{+18}} with the largest amount of fluorine, is less energetically favourable.

\begin{figure*}[th!]
    \centering
    \includegraphics[width=0.49\textwidth]{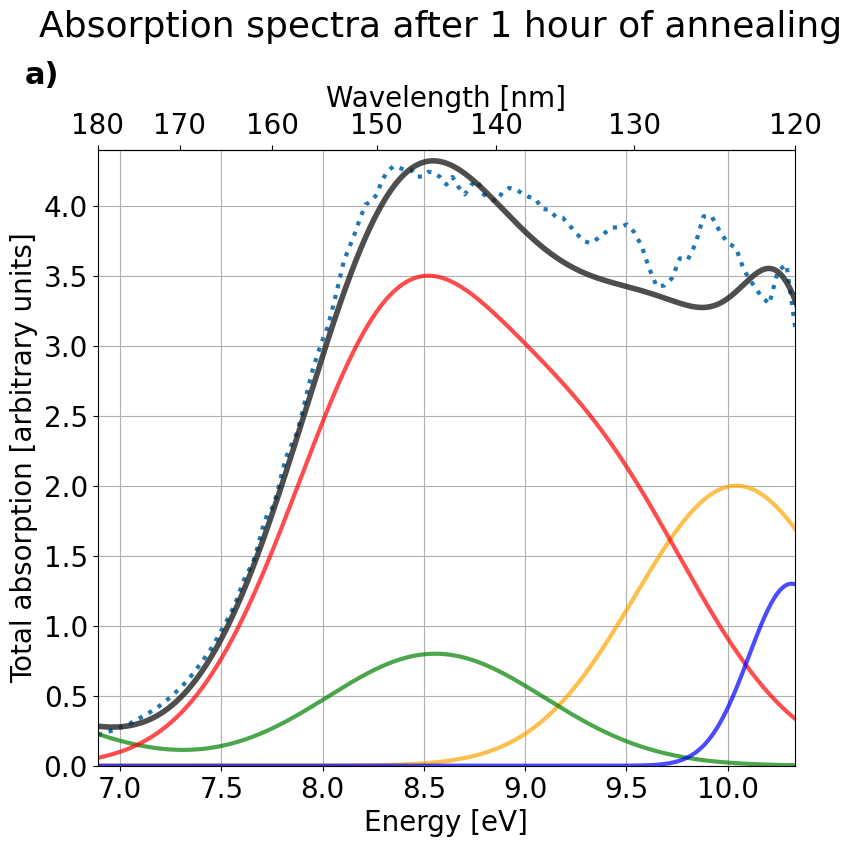}
    \includegraphics[width=0.49\textwidth]{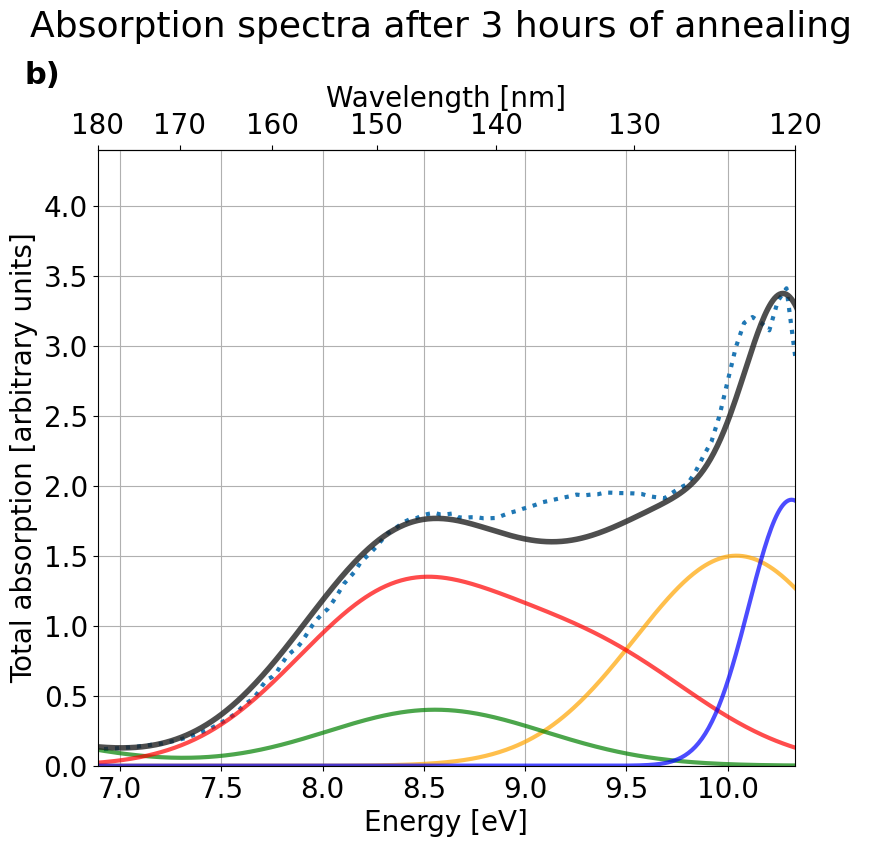}
    \includegraphics[width=0.49\textwidth]{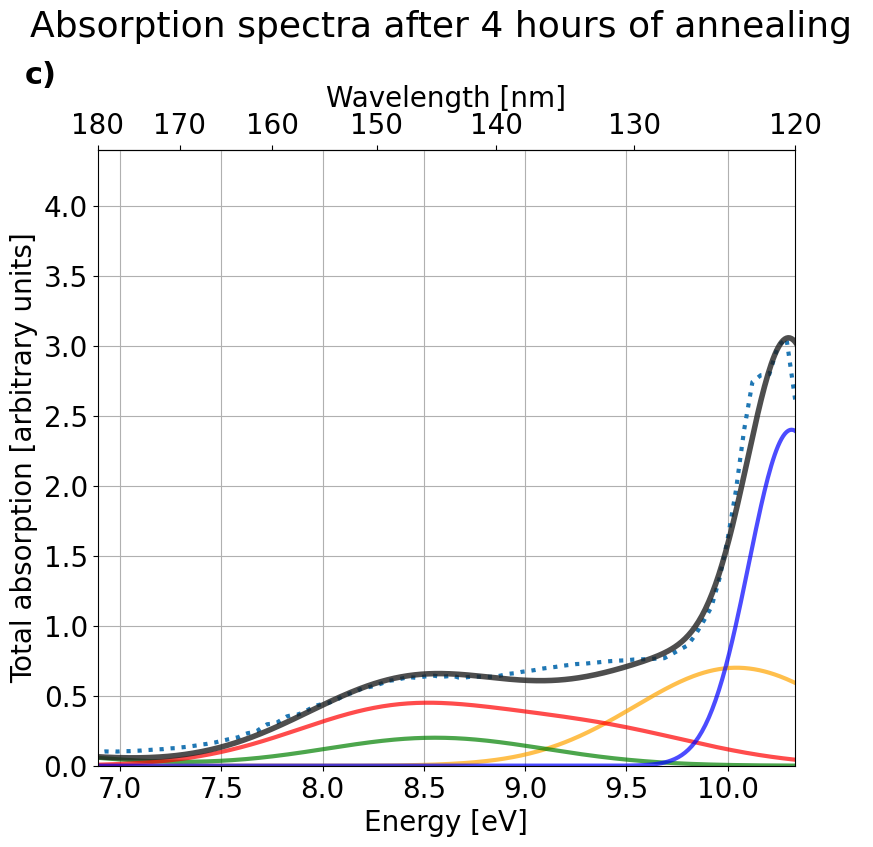}
    \includegraphics[width=0.49\textwidth]{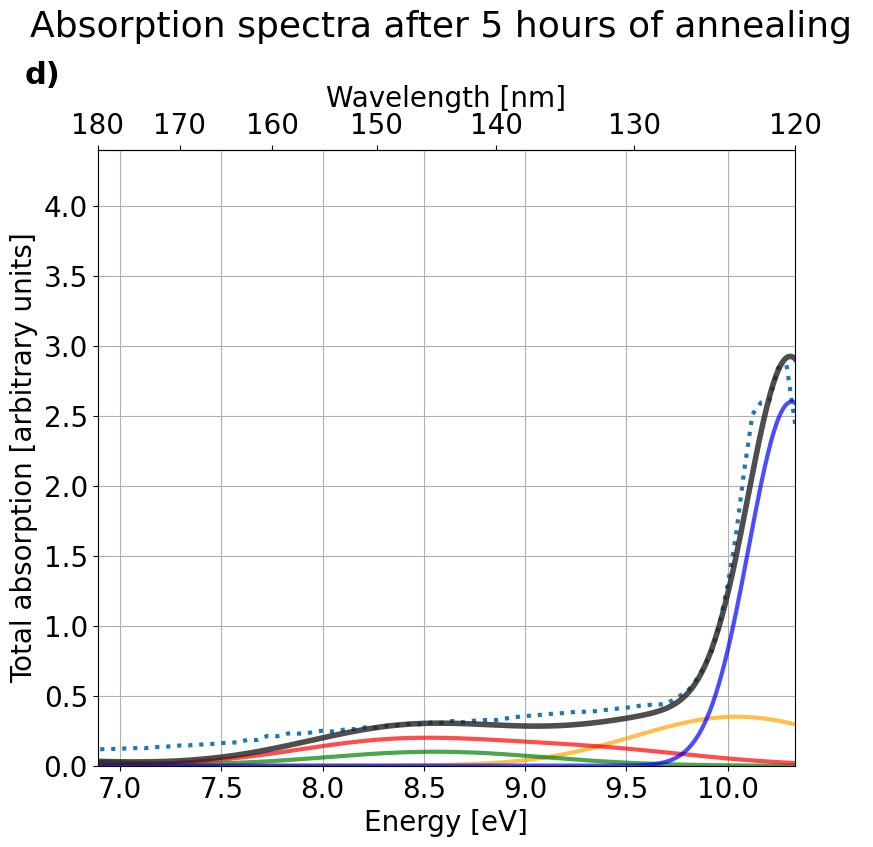}

    \caption{Comparison of simulated and experimental absorption spectra for different types of defect structures in the material. The dashed line represents experimental data after 1, 3, 4, and 5 hours of annealing, respectively. The colored lines indicate contributions from different types of defect structures: calcium vacancy (orange), double fluorine ion vacancy (green), single fluorine ion vacancy (red), uncompensated substitutional site (blue), and the total simulation (gray).}
    \label{fig:absfit}
\end{figure*}

It was shown that, due to the growth process, a thorium doped \ce{CaF_2} crystal is originally in a fluorine deficient state~\cite{beeks2023growth} with inherently many fluorine vacancies, resulting in reduced VUV transmission or even optically opaque samples. To re-gain the VUV transmission, the crystals are heated to a superionic state in a fluorinating atmosphere to replenish the fluorine~\cite{beeks2024optical}. 
Figure~\ref{fig:absfit} shows absorption spectra recorded after 1h, 3h, 4h, and 5h of fluorine annealing for the $^{229}$Th:CaF$_2$ crystal "X2", used in the recent nuclear laser spectroscopy and x-ray excitation experiments~\cite{tiedau2024laser,hiraki2024controlling,zhang2024frequency,higgins2024temperature}.
 
To model the annealing of a fluorine deficient crystal, we considered several models of defect structures (see table~\ref{tab:properties}) with an increasing number of fluorine ions around the thorium dopant. In Figure~\ref{fig:absfit}, 
we present the ab initio absorption spectra of the thorium site deficient in two \ce{F^-} (cluster-\ce{[ThF_6Ca_{12}]^{+22}}, green line), one \ce{F^-}  (cluster- \ce{[ThF_7Ca_{12}]^{+21}}, red line) as well as uncompensated thorium in perfect \ce{CaF_2} (cluster- \ce{[ThF_8Ca_{12}]^{+21}}, blue line) and  thorium site compensated by a calcium vacancy (\ce{[ThF_8Ca_{11}]^{+18}}, orange line). 
The time dependence of the absorption profile of the increasingly F-annealed sample can be explained by considering a time-dependent concentration of different types of defects. In Figure~\ref{fig:absfit},  we present the combined absorption spectra of aforementioned sites (gray solid line). The absorption profiles of individual clusters, along with their respective weights and Gaussian broadening factors, can be found in the Supplemental Material file \cite{supp}.

The absorption spectra of the 1h and 3h annealed sample show significant contribution from the one \ce{F-} and two \ce{F-} deficient sites, centred at 8.5\,eV and 8.5--9.5\,eV, respectively (panel a), b), red and green line).  In contrast, absorption in the 4h and 5h annealed crystal around 10.0\,eV  (panel c), d), orange line) originates primarily from four valent thorium at a calcium vacancy compensated site and thorium coordinated with 8 \ce{F-}. 

The calcium vacancy defect configurations yields absorption lines and hence electronic excited levels around 10\,eV, which are between the isomer excitation energy of 8.4\,eV and the \ce{CaF_2} direct band gap of 11.8\,eV. For both configurations, uncompensated substitutional \ce{Th} and \ce{Ca} vacancy, the one electron density of the excited states exhibits a high probability of finding an electron near the nucleus (see Figure~\ref{fig:vCadiff}).

The first excited electronic state of thorium  \ce{[ThF_{10}Ca_{12}]^{+18}}, fully compensated by two interstitial \ce{F^-} in the geometry optimized by Pimon~\cite{pimon2022ab}, lies above the band gap, and therefore this site is optically inactive, meaning its presence cannot be confirmed by optical absorption spectroscopy. 

In an experiment recently reported in~\cite{hiraki2024controlling}, a segment of the X2 crystal was irradiated with high energy x-rays, reducing the isomer half-live by up to a factor~50. The absorption of the one \ce{F-} and two \ce{F-} deficient defect geometries overlaps with the isomer excitation energy (8.4\,eV), making it plausible that fluorine deficient centers created upon high energy x-ray irradiation are responsible for opening a quenching channel for the isomer.

Electronic states within the bandgap can be exploited not only to quench the nucleus, but also to excite it more efficiently. The presence of a defect state in the perfect, stoichiometric crystal at 10\,eV provides a controllable platform to efficiently excite and quench the nucleus. As described in~\cite{nickerson2020nuclear,nickerson2021driven}, by exciting the defect and using a second laser to couple the defect with the nucleus, "bridging" the gap in energy, the interaction strength can be increased. Using this approach, the nucleus can be excited a hundred times more efficiently compared to direct photon excitation of the nucleus. Another crucial factor is the overlap of the electronic and nuclear wavefunction, along with non-zero transition matrix elements. As the excitation of the thorium impurity accompanied by the Ca vacancy with perfect 8 \ce{F-} coordinated sites~(see Figure \ref{fig:vCadiff}) transfers charge from the F\textsuperscript{-} to the Th, it is highly likely that this defect can be used to interact with the nucleus. The bridging laser can be used as an on/off switch for exciting or quenching the nucleus, which would allow for better clock operation~\cite{nickerson2021driven} and possibly creating a nuclear laser~\cite{tkalya2011proposal}.

The cluster method also allows for a flexible adjustment of the number of electrons in the Quantum Part. We have considered electronic structure properties of the clusters described above, "charged" with one or two additional electrons. Calculations showed that thorium becomes three or two valent, respectively. Moreover, in both cases, the excited state energy levels absorb light in the visible region. Given that the X2 sample is optically transparent when fully fluorinated~\cite{beeks2024optical}, we can conclude that during fluorine annealing, \ce{F-} ions move within the lattice to compensate fluorine vacancies, and thorium in \ce{CaF_2} in the electronic ground state is four valent.

\textit{Conclusions}.--
As described in~\cite{nickerson2020nuclear,nickerson2021driven} and confirmed by recent x-ray irradiation~\cite{hiraki2024controlling} and implantation experiments~\cite{pineda2024radiative}, defect states that are energy resonant with the isomer energy strongly influence the half-life of the $^{229}$Th isomer. Our ab initio absorption spectra of one \ce{F-} and two \ce{F-} deficient defect geometries confirm the presence of local electronic states within the energy range of the isomer excitation. By combining the calculated ab initio absorption spectra of individual active centers, we modeled the evolution of the observed absorption spectra in \textsuperscript{229}Th:CaF\textsubscript{2} during the annealing process, progressively reducing fluoride deficiencies. 

Further research is needed to determine if the quenching factors of the single-center absorption intensities can be quantitatively related to the concentration of the centers in the crystal lattice. Moreover, local multi-reference cluster methods provide electronic wave function of the ground and excited states which will be used in further research to calculate EB couplings between the electronic states and the nucleus.

\textit{Acknowledgements}--
We thank Martin Pimon for fruitful discussions and for providing optimized geometries of the sites \ce{[ThF_10Ca_12]^{+18}} and  \ce{[ThF_8Ca_11]^{+18}}.
MK, KN and VV thanks for the support to  Polish National Agency
for Academic Exchange under the Strategic Partnership Programme grant BNI/PST/2023/1/00013/U/00001.  VV thanks computer resources provided by the Swedish National Infrastructure for Computing (NAISS) at LUNARC.
KB acknowledges support from the Schweizerischer Nationalfonds (SNF), fund 514788 “Wavefunction engineering for controlled nuclear decays".
This work has been funded by the European Research Council (ERC) under the European Union’s Horizon 2020 research and innovation programme (Grant Agreement No. 856415) and the Austrian Science Fund (FWF) [Grant DOI: 10.55776/F1004, 10.55776/J4834, 10.55776/ PIN9526523]. The project 23FUN03 HIOC [Grant DOI: 10.13039/100019599] has received funding from the European Partnership on Metrology, co-financed from the European Union’s Horizon Europe Research and Innovation Program and by the Participating States.

\bibliography{biblio}
\bibliographystyle{ieeetr}
\end{document}


\preprint{APS/123-QED}

\title{An Embedding Cluster Approach for Accurate Electronic Structure Calculations of (229)Th:\ce{CaF2}.\\
	Supplemental material}

\author{Kamil Nalikowski$^{1}$}
\author{Valera Veryazov$^{2}$}
\author{Kjeld Beeks$^{3,4}$}
\author{Thorsten Schumm$^{3}$}
\email{email: thorsten.schumm@tuwien.ac.at}
\author{Marek Kro\'{s}nicki$^{1}$}
\email{email: marek.krosnicki@ug.edu.pl}

\affiliation{$^1$Institute of Theoretical Physics and Astrophysics, Faculty of Mathematics Physics and Informatics, University of Gda\'{n}sk, ul. Wita Stwosza 57, Gda\'{n}sk, 80-308, Poland}
\affiliation{$^2$Division of Computational Chemistry, Chemical Center, Lund University, P.O. Box 124, SE-221 00 Lund, Sweden}
\affiliation{$^3$Vienna Center for Quantum Science and Technology, Atominstitut, TU Wien, 1020 Vienna, Austria}
\affiliation{$^4$Laboratory for Ultrafast Microscopy and Electron Scattering (LUMES), Institute of Physics, École Polytechnique Fédérale de Lausanne (EPFL), Lausanne CH-1015, Switzerland}

\maketitle

\section{Computational details}
\subsection{Coordinates}

Cartesian coordinates of $[ThF_{10}Ca_{12}]$ cluster in Angstrom.

\begin{verbatim}

    Th1       0.000000       0.000000        0.000000
    F1       -0.392304       0.392025        2.235551  
    F2       -0.392307       2.235546        0.392033  
    F3       -1.161103      -1.789561        1.161330  
    F4        1.488018       1.268511       -1.487791 
    F5        1.789363       1.161656        1.161659  
    F6       -1.268273      -1.488035       -1.488030 
    F7        1.424356      -1.424111       -1.424106 
    F8       -2.235825       0.391281        0.391286  
    F9       -1.161100       1.161324       -1.789556 
    F10       1.488019      -1.487796        1.268517 
    Ca1      -2.732675      -2.896178       -0.082478 
    Ca2      -2.707109       2.707254       -0.069414 
    Ca3       2.877466      -2.877160       -0.068286 
    Ca4       2.896348       2.732923       -0.082310 
    Ca5      -2.732674      -0.082479       -2.896175 
    Ca6      -2.707111      -0.069416        2.707258  
    Ca7       2.877464      -0.068287       -2.877157 
    Ca8       2.896349      -0.082311        2.732927  
    Ca9       0.068579      -2.877283       -2.877282 
    Ca10      0.082630      -2.896164        2.732979 
    Ca11      0.082629       2.732976       -2.896161
    Ca12      0.069540       2.707230        2.707231 

\end{verbatim}

Cartesian coordinates of $[ThF_{8}Ca_{11}]$ cluster in Angstrom.

\begin{verbatim}
    Th1       0.000000       0.000000        0.000000   
    F1       -1.200224       1.494927       -1.200225  
    F2       -1.200224      -1.494927       -1.200225  
    F3       -1.297665      -1.329964        1.433859  
    F4       -1.297665       1.329963        1.433859  
    F5        1.380710       1.341197        1.380712  
    F6        1.380710      -1.341197        1.380712  
    F7        1.433859      -1.329964       -1.297665  
    F8        1.433859       1.329963       -1.297665  
    Ca1      -2.823389      -2.831236        0.007685  
    Ca2      -2.823389       2.831237        0.007685  
    Ca3       2.874449      -2.834242        0.032164  
    Ca4       2.874449       2.834243        0.032164  
    Ca5      -2.801793       0.000000        2.898511  
    Ca6       2.898512       0.000000       -2.801794  
    Ca7       2.875302       0.000000        2.875301  
    Ca8       0.007684      -2.831236       -2.823388  
    Ca9       0.032164      -2.834242        2.874447  
    Ca10      0.007684       2.831237       -2.823388  
    Ca11      0.032164       2.834243        2.874447  
\end{verbatim}
\subsection{Basis set and ab initio model potentials}
\begin{table*}[h!]
    \centering
    \begin{tabular}{ccc}
    \toprule
         & atom & \\
               Th & F & Ca \\ \midrule
          ANO-RCC...9s8p6d4f2g1h & ANO-XS...3s2p1d. & ANO-XS...5s3p.\\
    \bottomrule
    \end{tabular}
    \caption{Basis set used in every performed cluster calculation.  }
    \label{tab:basis_set}
\end{table*}

\subsection{Active space selection}
\begin{table*}[h!]
    \centering
    \begin{tabular}{lcclll}
    \toprule \midrule
    Quantum part & Active orbitals & Active electrons \\ \midrule
    \ce{[ThF_6Ca_12]^{+22}}& 9 & 6\\
    \ce{[ThF_7Ca_12]^{+21}}& 12 & 10\\
         \ce{[ThF_8Ca_11]^{+18}}& 14 & 8 \\
         \ce{[ThF_8Ca_12]^{+20}}& 10 & 10\\
        \ce{[ThF_10Ca_12]^{+18}}& 13 & 12\\         
         \midrule \bottomrule
    \end{tabular}
    \caption{Size of an active space of each quantum part.}
    \label{tab:active space}
\end{table*}

\begin{figure}[h!]
    \centering
    \includegraphics[width=0.99\linewidth]{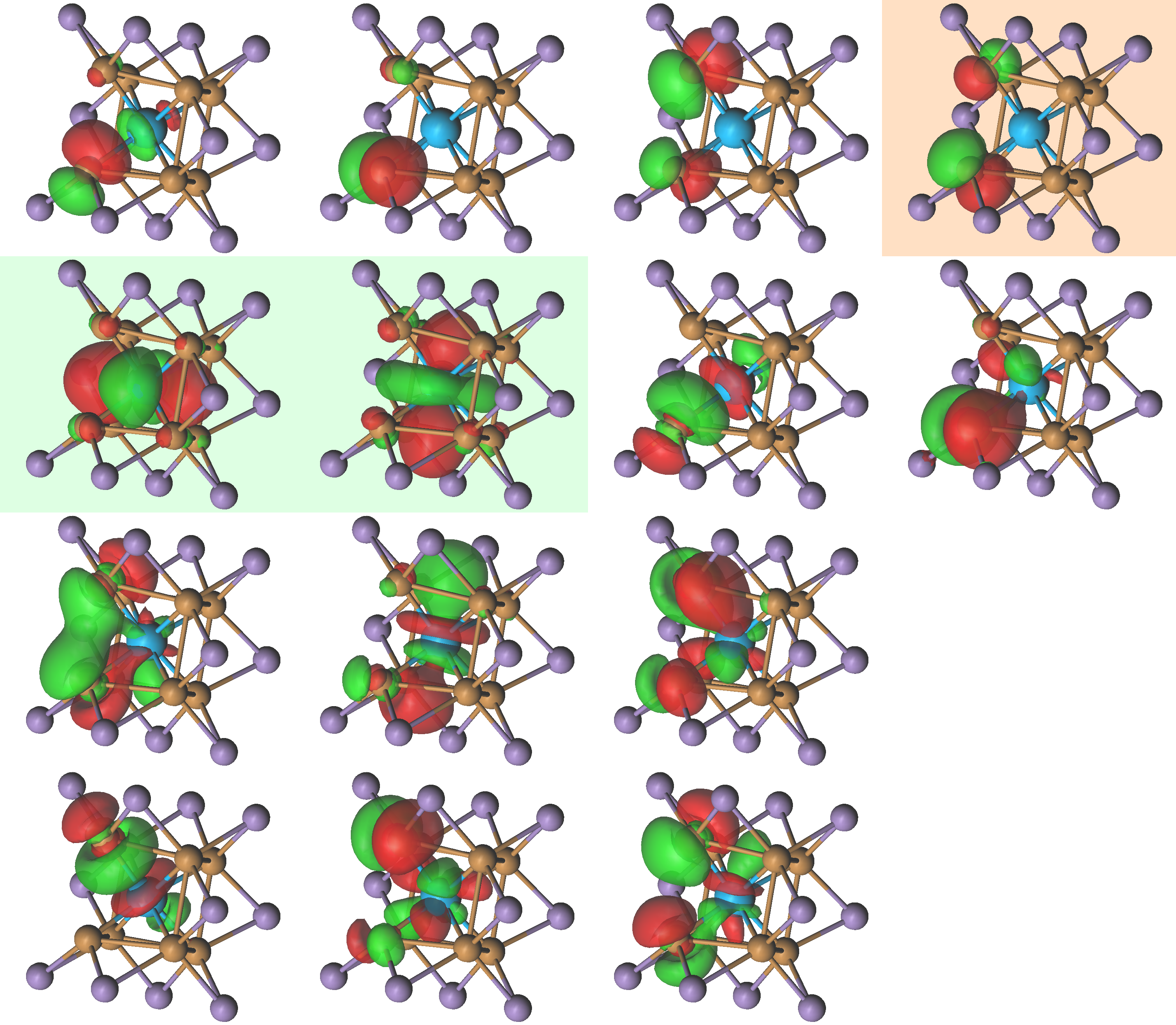}
    \caption{Active space for \ce{ThF_{8}Ca_{11}} cluster. Orange background tile is the molecular orbital from which electrons are mostly taken; while green background tiles correspond to molecular orbitals which electrons are mostly excited on.}
    \label{fig:activespace}
\end{figure}

\newpage

\section{Atomic charges and valence of $Th$ in $ThF_n$ molecules}
\begin{table}[h!]
    \centering
    \begin{tabular}{|c|c|c|c|}
    \hline
         Molecule & $Th$ Atomic Charge & $Th$ Valence  & $Th-F$ bond order\\
         \hline
         \ce{ThF}& 0.69 & 1.14 & 0.72\\
         \ce{ThF_2}& 1.37 & 2.23 & 0.70\\
         \ce{ThF_3}& 2.08 & 3.18 & 0.63\\
         \ce{ThF_4}& 2.82 & 4.12 & 0.54\\
         \hline
    \end{tabular}
    \caption{Atomic charges and valence of $Th$ and a bond order of $Th-F$ bond in molecular fluorides of thorium.}
    \label{tab:my_label}
\end{table}

\section{Total energy of computed clusters}

\begin{table*}[h!]
    \centering
    \begin{tabular}{|l|r|}
    \toprule \midrule
    Quantum part &  Total Energy a.u.\\
         \hline
         \ce{[ThF_6Ca_12]^{+22}}& -35194.072\\
         \ce{[ThF_7Ca_12]^{+21}}&-35294.588\\
         \ce{[ThF_8Ca_11]^{+18}}&-34716.031\\
         \ce{[ThF_8Ca_12]^{+20}}& -35394.977\\
        \ce{[ThF_10Ca_12]^{+18}}&-35594.857\\
         \midrule \bottomrule
    \end{tabular}

    \caption{
    The CASPT2 energy of the computed clusters}
    \label{tab:properties}
\end{table*}

\section{Absorption data}

\begin{table}[h!]
    \centering
    \begin{tabular}{ccc}
    \toprule \midrule
         index & Energy [eV] & Osc. Str. [au] \\ \midrule
         1& 0.000 & --\\
        2 & 9.719 & 1.980E-08 \\
        3 & 9.719 & 5.485E-05 \\
        4 & 9.723 & 3.773E-10 \\
        5 & 10.019 & 1.203E-04 \\
        6 & 10.041 & 3.342E-02 \\
        7 & 10.054 & 1.478E-04 \\
        8 & 10.058 & 3.771E-08 \\
        9 & 10.059 & 1.967E-03 \\
        10 & 10.290& $<$ 1E-10 \\ 
        11 & 10.292 & 3.004E-04 \\
        12 & 10.295 & 2.466E-06 \\
        \midrule \bottomrule
    \end{tabular}
    \caption{Calculated excitation energies of different electronic state and their  oscillator strength [au] in the case of $\ce{ThF8C11}$}
    \label{tab:vCatable}
\end{table}

\begin{figure}[h!]
    \centering
    \includegraphics[width=0.5\linewidth]{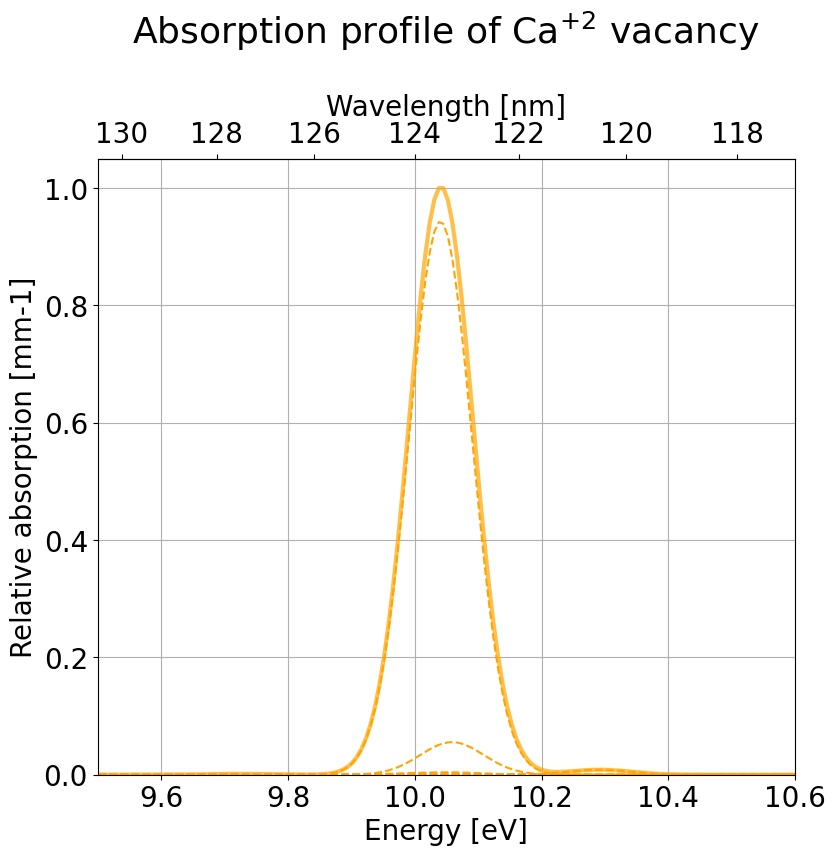}
    \caption{Total absorption profile (solid line) accompanied with a normalized gaussian profiles of standard deviation equal to $0.05$ eV, mean values and heights given by excitation energies and weights from table \ref{tab:vCatable}, respectively.}
    \label{fig:sivCa}
\end{figure}

\begin{table}[]
    \centering
    \begin{tabular}{ccc}
    \toprule \midrule
         index & Energy [eV] & Osc. Str. [au] \\ \midrule
         1 & 0.000 & --\\
        2 & 6.185 & 2.207E-03 \\
3 & 6.188 & 2.651E-03 \\
4 & 6.192 & 1.062E-03 \\
5 & 6.200 & 1.510E-03 \\
6 & 6.382 & 9.131E-04 \\
7 & 6.390 & 9.397E-05 \\
8 & 6.399 & 1.594E-03 \\
9 & 6.442 & 5.860E-04 \\
10 & 6.630 & 8.385E-05 \\
11 & 6.652 & $<$ 1E-05 \\
12 & 6.656 & 3.645E-05 \\
13 & 6.702 & 3.807E-04 \\
14 & 8.300 & 2.771E-03 \\
15 & 8.315 & 1.786E-04 \\
16 & 8.318 & 6.825E-05 \\
17 & 8.333 & 1.214E-03 \\
18 & 8.370 & 1.355E-04 \\
19 & 8.381 & 1.694E-05 \\
20 & 8.393 & 1.121E-03 \\
21 & 8.431 & 7.748E-04 \\
22 & 8.484 & 4.354E-03 \\
23 & 8.550 & 1.360E-04 \\
24 & 8.553 & 2.149E-04 \\
25 & 8.571 & 7.742E-04 \\
26 & 8.607 & 3.815E-04 \\
27 & 8.611 & 8.759E-04 \\
28 & 8.637 & 9.826E-05 \\
29 & 8.664 & 1.868E-03 \\
30 & 8.723 & 1.364E-03 \\
31 & 8.820 & 2.081E-03 \\
32 & 8.846 & 9.624E-05 \\
33 & 8.859 & 1.927E-04 \\
34 & 8.887 & 5.397E-04 \\
35 & 8.899 & 6.747E-05 \\
36 & 8.909 & 3.860E-05 \\
37 & 8.925 & 1.077E-03 \\
        \midrule \bottomrule
    \end{tabular}
    \caption{Calculated excitation energies of different electronic state and their  oscillator strength [au] in the case of $\ce{ThF6C12}$}
    \label{tab:v2Ftable}
\end{table}
\begin{figure}[h!]
    \centering
    \includegraphics[width=0.35\linewidth]{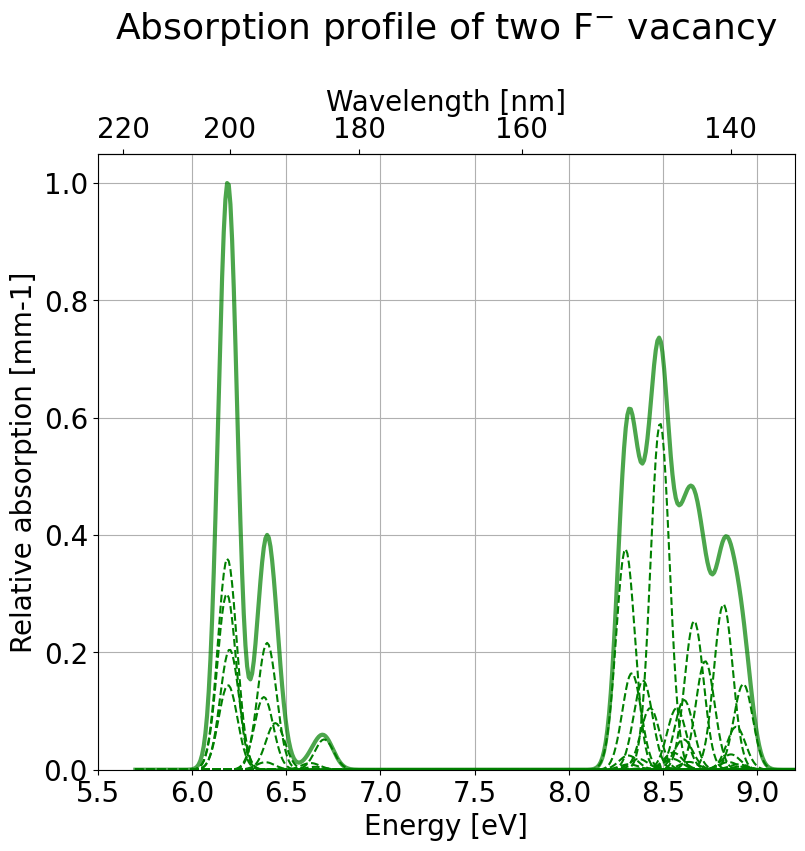}
    \caption{Total absorption profile (solid line) accompanied with a normalized gaussian profiles of standard deviations equal to $0.05$ eV, mean values and heights given by excitation energies and  oscillator strength  from table \ref{tab:v2Ftable}, respectively.}
    \label{fig:si2f}
\end{figure}
\clearpage
\newpage
\begin{table}[]
    \centering
    \begin{tabular}{ccc}
    \toprule \midrule
         index & Energy [eV] & Osc. Str.[au] \\ \midrule
         1 & 0.000 & -- \\
2 & 8.027 & 1.540E-03 \\
3 & 8.074 & 3.162E-04 \\
4 & 8.080 & 2.070E-04 \\
5 & 8.084 & 2.779E-04 \\
6 & 8.200 & 2.625E-03 \\
7 & 8.272 & 9.791E-05 \\
8 & 8.277 & 2.349E-04 \\
9 & 8.279 & 9.392E-05 \\
10 & 8.366 & 5.584E-04 \\
11 & 8.383 & 1.238E-03 \\
12 & 8.385 & 2.303E-03 \\
13 & 8.386 & 5.776E-05 \\
14 & 8.413 & 6.696E-03 \\
15 & 8.437 & 9.278E-04 \\
16 & 8.438 & 1.107E-03 \\
17 & 8.440 & 5.145E-04 \\
18 & 8.476 & 2.746E-03 \\
19 & 8.494 & 1.696E-03 \\
20 & 8.494 & 2.616E-04 \\
21 & 8.500 & 6.066E-04 \\
22 & 9.262 & 7.205E-03 \\
23 & 9.367 & 6.189E-03 \\
24 & 9.713 & 1.775E-03 \\
    \midrule \bottomrule
    \end{tabular}
    \caption{Calculated excitation energies of different electronic state and their  oscillator strength [au] in the case of $\ce{ThF7C12}$}
    \label{tab:v1Ftable}
\end{table}

\begin{figure}[h!]
    \centering
    \includegraphics[width=0.5\linewidth]{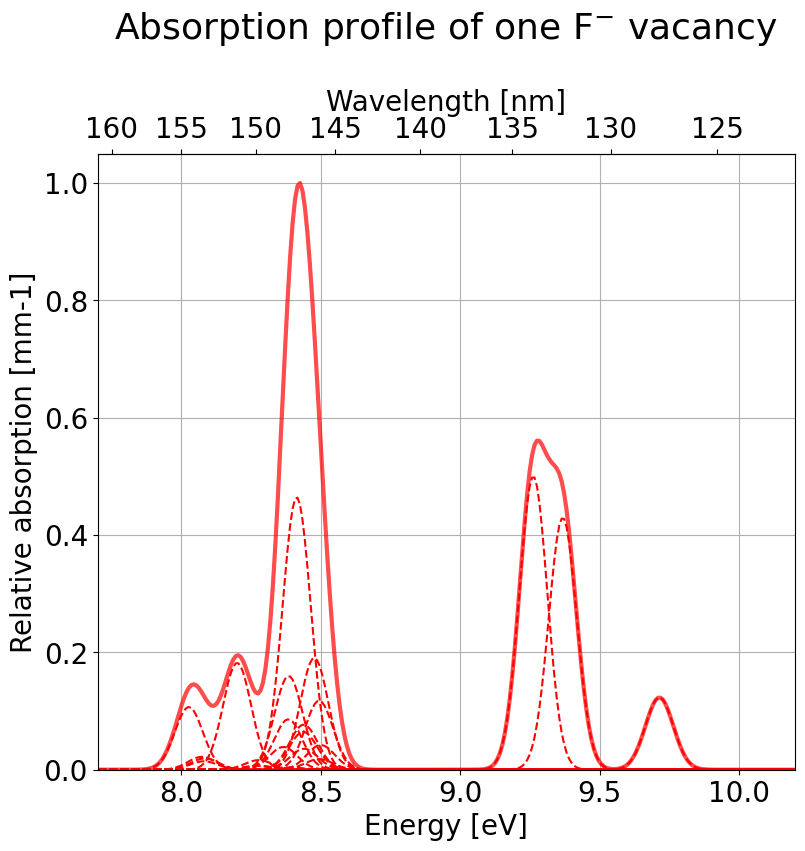}
    \caption{Total absorption profile (solid line) accompanied with a normalized gaussian profiles of standard deviations equal to $0.05$ eV, mean values and heights given by excitation energies and  oscillator strength from table \ref{tab:v1Ftable}, respectively.}
    \label{fig:si1f}
\end{figure}

\begin{table}[]
    \centering
    \begin{tabular}{ccc}
    \toprule \midrule
         index & Energy [eV] & Osc. Str. [au]\\ \midrule
1 & 0.000 & --\\
2 & 10.284 & 6.043E-04 \\
3 & 10.395 & 1.028E-06 \\
4 & 10.396 & 1.459E-04 \\
5 & 10.403 & 2.416E-05 \\
6 & 10.667 & 2.069E-08 \\
7 & 10.671 & 8.032E-05 \\
8 & 10.671 & 1.281E-06 \\
9 & 11.383 & 9.368E-10 \\
10 & 11.513 & 1.738E-07 \\
11 & 11.513 & 3.724E-08 \\
12 & 11.513 & 9.689E-06 \\
    \midrule \bottomrule
    \end{tabular}
    \caption{Calculated excitation energies of different electronic state and their  oscillator strength [au] in the case of $\ce{ThF8C12}$}
    \label{tab:vpuretable}
\end{table}
\begin{figure}[h!]
    \centering
    \includegraphics[width=0.5\linewidth]{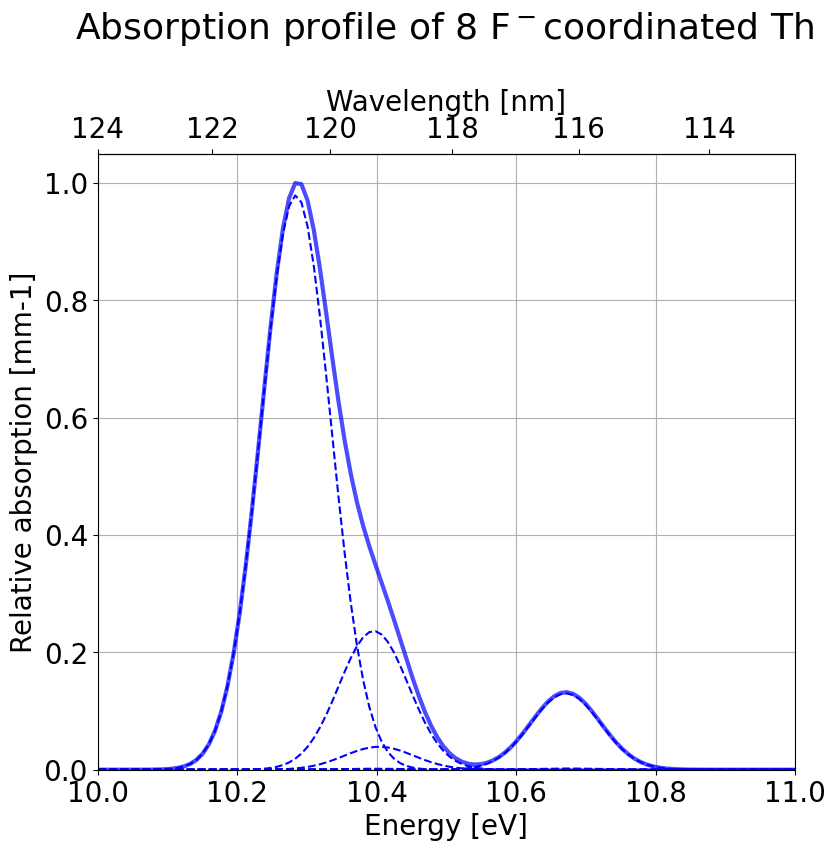}
    \caption{Total absorption profile (solid line) accompanied with a normalized gaussian profiles of standard deviations equal to $0.05$ eV, mean values and heights given by excitation energies and  oscillator strength  from table \ref{tab:vpuretable}, respectively.}
    \label{fig:sipure}
\end{figure}

\clearpage
Fitting goes as follow: for each excitation form a gaussian curve centered on calculated energy with a height given by oscillator strength and fitted broadening ($\sigma$). Then sum the all the cotribution coming from each quantum part and normalize the highest peak to 1. The total simultion spectra comes from weighted sum of each individual spectra with a weight given by scale - see table \ref{tab:fitparam}.

\begin{table}[ht]
\centering
\begin{tabular}{lcccc}
\toprule
\textbf{Quantum cluster} & \textbf{vCa} & \textbf{ two F-} & \textbf{one F-} & \textbf{Pure} \\
\midrule
\textbf{$\sigma$}       & 0.5   & 0.5   & 0.5   & 0.2   \\
\textbf{Scale (1h)}  & 2.00  & 0.80  & 3.50  & 1.30  \\
\textbf{Scale (3h)}  & 1.50  & 0.40  & 1.35  & 1.90  \\
\textbf{Scale (4h)}  & 0.70  & 0.20  & 0.45  & 2.40  \\
\textbf{Scale (5h)}  & 0.35  & 0.10  & 0.20  & 2.60  \\
\bottomrule
\end{tabular}
\caption{Fit parameters for different quantum clusters and annealing times.} \label{tab:fitparam}
\end{table}